\documentclass[aps,pra,groupedaddress,floatfix,nofootinbib]{revtex4}

\usepackage{amsmath}
\usepackage{graphicx}
\usepackage{color}

\begin{document}

\title{On the variational homotopy perturbation method for nonlinear oscillators}

\author{Francisco M. Fern\'andez}
\email{fernande@quimica.unlp.edu.ar}

\affiliation{INIFTA (UNLP, CONICET), Divisi\'{o}n Qu\'{i}mica
Te\'{o}rica, Diag. 113 y 64 (S/N), Sucursal 4, Casilla de Correo
16, 1900 La Plata, Argentina}

\date{\today}

\begin{abstract}
In this paper we discuss a recent  application of a variational homotopy
perturbation method to rather simple nonlinear oscillators . We show that
the main equations are inconsistent and for that reason the results may be
of scarce utility.
\end{abstract}

\maketitle

\section{Introduction}

There has recently been great interest in developing simple
solutions to textbook models of nonlinear oscillators\cite{H03,
RH09,H10,AL11} (and references therein). However, some of them are
of questionable utility as shown, for example, by Rajendran et
al\cite{RPL04} who concluded that He's calculations of the limit
cycle of the van der Pol oscillator\cite{H03} ``contain several
errors which once rectified make the method inapplicable to it''.
I have disclosed several inconsistencies in a paper by Ren and He
\cite{RH09} and even proposed how to tidy up and improve their
calculations \cite{F09a}.

Here I discuss a recent application of a variational homotopy perturbation
method to rather simple nonlinear oscillators\cite{AL11}. In Sec.~\ref
{sec:Variational_homotopy} I analyse their results and in Sec.~\ref
{sec:Conclusions} draw conclusions.

\section{Variational homotopy perturbation method for nonlinear oscillators}

\label{sec:Variational_homotopy}

Akbarzade and Langari\cite{AL11} were interested in equations of the form
\begin{equation}
A(u)-f(r)=L(u)+N(u)-f(r)=0  \label{eq:LN}
\end{equation}
where $L$ and $N$ are the linear and nonlinear parts of the
operator $A$ and $u$ is the solution. They proposed the ``homotopy
perturbation structure''
\begin{equation}
H(v,p)=(1-p)[L(v)-L(u_{0})]+p[A(v)-f(r)]=0  \label{eq:H(v.p)}
\end{equation}
where $p$ is an embedding parameter (dummy perturbation parameter in the
language of the well known perturbation theory) and $u_{0}$ is the first
approximation that satisfies the boundary conditions.

They expanded the solution in $p$--power series
$v=v_{0}+v_{1}p+v_{2}p^{2}+ \ldots $ and obtained the solution to
Eq.~(\ref{eq:LN}) as $ u=v_{0}+v_{1}+v_{2}+\ldots $ provided that
the series converges for $p=1$.

In particular, the authors concentrated in nonlinear oscillators of the form
\begin{equation}
u^{\prime \prime }+\omega _{0}^{2}u+\epsilon f(u)=0  \label{eq:gen_osc}
\end{equation}
where $f$ is a nonlinear function of $u^{\prime \prime }$, $u^{\prime }$ and
$u$, and considered the ``variational functional''\cite{AL11} (and
references therein)
\begin{equation}
J(u)=\int_{0}^{t}\left[ -\frac{1}{2}u^{\prime 2}+\frac{1}{2}\omega
_{0}^{2}u^{2}+\epsilon F(u)\right] dt  \label{eq:var_fun}
\end{equation}
where $dF/du=f$. Note that I have corrected a misprint in the authors'
Eq.~(9). Obviously, $J(u)$ is minus the well known action integral\cite{G80}
for a particular time interval.

In order to introduce the basic idea the authors first modified the
well--known Duffing equation
\begin{equation}
u^{\prime \prime }+u+\epsilon u^{3}=0,\;u(0)=A,\;u^{\prime }(0)=0
\label{eq:Duffing}
\end{equation}
as
\begin{equation}
u^{\prime \prime }+\omega ^{2}u+p\left[ \epsilon u^{3}+(1-\omega
^{2})u\right] =0  \label{eq:Duffing_homo}
\end{equation}
and derived the perturbation equations of order cero
\begin{equation}
u_{0}^{\prime \prime }+\omega ^{2}u_{0}=0  \label{eq:Duffing_0}
\end{equation}
and first order
\begin{equation}
u_{1}^{\prime \prime }+\omega ^{2}u_{1}+\epsilon u_{0}^{3}+(1-\omega
^{2})u_{0}=0  \label{eq:Duffing_1}
\end{equation}
where
\begin{equation}
u_{0}(t)=A\cos (\omega t)  \label{eq:u0}
\end{equation}
satisfies the boundary conditions and
\begin{equation}
u_{1}(0)=u_{1}^{\prime }(0)=0  \label{eq:u1_BC}
\end{equation}

According the the authors ``$\omega $ will be identified from the
variational formulation for $u_{1}$, which reads''
\begin{equation}
J(u_{1})=\int_{0}^{T}\left[ -\frac{1}{2}u_{1}^{\prime 2}+\frac{1}{2}\omega
^{2}u_{1}^{2}+(1-\omega ^{2})u_{0}u_{1}+\epsilon u_{0}^{3}u_{1})\right]
dt,\;T=\frac{2\pi }{\omega }  \label{eq:Duffing_J(u1)}
\end{equation}
They argued that the simplest trial function is\cite{AL11}
\begin{equation}
u_{1}=B\left[ \cos (\omega t)-\frac{1}{3}\cos (5\omega t)\right]
\label{eq:Dufing_u1_a}
\end{equation}
Surprisingly, this function satisfies one of the boundary
conditions $ u_{1}^{\prime }(0)=0$ but not the other one because
$u_{1}(0)=2B/3=0$ leads to the unwanted trivial solution.

From the variational conditions $\partial J/\partial B=0$ and $\partial
J/\partial \omega =0$ the authors obtained
\begin{equation}
\omega =\sqrt{1+\frac{3}{4}\epsilon A^{2}},\;B=0
\label{eq:Duffing_Omega_B_AL}
\end{equation}
Although the estimated value of $\omega $ is reasonable, the result $B=0$
leads to the trivial solution $u_{1}\equiv 0$ that restricts considerably
the practical utility of the approach.

In order to improve the results the authors proposed the correction

\begin{equation}
u_{1}=B_{1}\left( \cos {\left( \omega t\right) }-\frac{\cos
{\left( 3\omega t\right) }}{5}\right) +B_{3}\left( \frac{\cos
{\left( 3\omega t\right) }}{5}- \frac{\cos {\left( 5\omega
t\right) }}{7}\right)  \label{eq:Dufing_u1_b}
\end{equation}
and from the variational conditions $\partial J/\partial B_{1}=0$, $\partial
J/\partial B_{3}=0$, $\partial J/\partial \omega =0$ they obtained the
frequency

\begin{equation}
\omega =\frac{\sqrt{31}}{124}\sqrt{\sqrt{510237\rho
^{2}+1416576\rho +984064} -357\rho -496}
\label{eq:Duffing_Omega_2}
\end{equation}
They did not show the coefficients; I obtained

\begin{equation}
B_{1}=\frac{A\left[ 357\rho -496\left( \omega ^{2}-1\right)
\right] }{ 96\omega ^{2}},\;B_{3}=\frac{49A\left[ 3\rho -4\left(
\omega ^{2}-1\right) \right] }{96\omega ^{2}}
\label{eq:Duffing_B1_B3}
\end{equation}
Note that $u_{1}(t)$ does not satisfy one of the required boundary conditions

\begin{equation}
u_{1}(0)=-\frac{A\left( 68\omega ^{2}-49\rho -68\right) }{16\omega ^{2}}
\end{equation}
and that the approximate solution $u_{app}(t)=u_{0}(t)+u_{1}(t)$ exhibits
the wrong amplitude $u_{app}(0)=A+u_{1}(0)$ . It therefore seems that by
means of the variational approach the authors obtained a frequency for the
amplitude $A$ and an approximate trajectory $u_{app}(t)$ with a different
amplitude.

Akbarzade and Langari\cite{AL11} applied the method to other
textbook nonlinear oscillators and obtained approximate
frequencies. In all the cases they chose first--order corrections
$u_{1}(t)$ that do not satisfy the boundary condition $u_{1}(0)=0$
and obtained the trivial solution $ u_{1}(t)\equiv 0$.

\section{Conclusions}

\label{sec:Conclusions}

Although the combination of the homotopy perturbation method and the
variational principle proposed by Akbarzade and Langari\cite{AL11} led to
reasonably approximate frequencies, one cannot take the approach seriously
because of its inconsistencies. First, the first--order corrections to the
solutions do not satisfy one of the boundary conditions proposed by the
authors and, second, in most of the cases the resulting corrections are
trivial (that is to say, they vanish identically). In the case where this
correction does not vanish, the approximate solution exhibits an amplitude
that is different from the one appearing in the expression for the frequency.


\begin{thebibliography}{9}
\bibitem{H03}  J-H He, Phys. Rev. Lett. \textbf{90}, 174301 (3 pp) (2003).

\bibitem{RH09}  Z-F Ren and J-H He, Phys. Lett. A \textbf{373}, 3749 (2009).

\bibitem{H10}  J-H He, Phys. Lett. A \textbf{374}, 2312 (2010).

\bibitem{AL11}  M. Akbarzade and J. Langari, J. Math. Phys. \textbf{52},
023518 (2011).

\bibitem{RPL04}  S. Rajendran, S . N. Pandey, and M. Lakshmanan, Phys. Rev.
Lett. \textbf{93}, 069401 (1 pp) (2004).

\bibitem{F09a}  F. M. Fern\'{a}ndez, On a simple approach to nonlinear
oscillators, arXiv:0910.0600v1 [math-ph]

\bibitem{G80}  H. Goldstein, Classical Mechanics, Second ed.
(Addison-Wesley, Reading, MA, 1980).
\end{thebibliography}
\end{document}